**Observation of Various and Spontaneous Magnetic Skyrmionic Bubbles at Room-Temperature in a Frustrated Kagome Magnet with Uniaxial Magnetic Anisotropy**

*Zhipeng Hou\*, Weijun Ren\*,Bei Ding\*, Guizhou Xu, Yue Wang, Bing Yang, Qiang Zhang, Ying Zhang,  Enke Liu, Feng Xu, Wenhong Wang,Guangheng Wu, Xi-xiang Zhang, Baogen Shen, Zhidong Zhang*

Dr. Z. P. Hou, B. Ding, Y. Wang, Dr. Y. Zhang, Dr. E. K. Liu, Prof. W. H. Wang, Prof. G. H. Wu, Prof. B. G. Shen
Beijing National Laboratory for Condensed Matter Physics, Institute of Physics, Chinese Academy of Sciences, Beijing 100190, China
E-mail: wenhong.wang@iphy.ac.cn (W. H. Wang)
Dr.W. J. Ren, Dr. B. Yang, Prof. Z. D. Zhang
Shenyang Materials Science National Laboratory, Institute of Metal Research, Chinese Academy of Sciences, 72 Wenhua Road, Shenyang 110016, China
E-mail:wjren@imr.ac.cn (W. J. Ren)
Dr. G. Z. Xu, Prof. F. Xu
School of Materials Science and Engineering, Nanjing University of Science and Technology, Nanjing 210094, China
Dr. Qiang Zhang, Prof. X. X. Zhang
King Abdullah University of Science and Technology (KAUST), Physical Science and Engineering (PSE), Thuwal 23955-6900, Saudi Arabia

[*] Z.P.H, W.J.R, and B.D contributed equally to this work.

Keywords: skyrmionic bubbles, topological spin textures, kagome magnet, $Fe_3Sn_2$, spintronic devices

The quest for materials hosting topologically protected nanometric spin textures, so-called magnetic skyrmions or magnetic skyrmionic bubbles, continues to be fuelled by the promise of novel devices.[1-5] The skyrmionic spin textures have been mostly observed in non-centrosymmetric crystals, such as the cubic chiral magnets MnSi,[6-9] (Mn,Fe)Ge,[10,11] FeCoSi,[12] $Cu_2OSeO_3$,[13-15] and also the polar magnet $GaV_4S_8$,[16] where Dzyaloshinskii-Moriya interaction (DMI) is active. A number of intriguing electromagnetic phenomena, including the topological Hall effect,[7] skyrmion magnetic resonance,[17] thermally induced ratchet motion,[18] and effective magnetic monopoles,[19] have been demonstrated to be closely related to the topologically nontrivial spin texture of skyrmions. These novel topological properties, together with nanoscale dimensions, a stable particle-like feature, and an ultralow threshold for current-driven motion, make magnetic skyrmions fundamentally promising for applications in next-generation high-density and low-dissipation memory devices.[1-5]

However, although a very recent publication by Tokunage *et al*.[20] reported the observation of skyrmion lattices above room temperature (RT) in *β*–type Cu-Zn-Mn alloys, skyrmions in the bulk chiral magnets have been mostly observed below RT.[6-16] Moreover, the spin texture of skyrmions that were stabilized by DMI in chiral magnets is quite limited. The skyrmions with variable spin textures may be more attractive for further technological applications because they can adapt to various external stimuli acting as information carriers in spintronic devices. Therefore, one particularly important current research direction aims at the discovery of new materials that host skyrmions with variable spin textures at room temperature.

In addition to the non-centrosymmetric chiral magnets in which the magnetic skyrmions are stabled by DMI, the centrosymmetric materials with uniaxial magnetic anisotropy (UMA) are another family of materials that can host skyrmions.[21-26] In these materials, the competition between the magnetic dipole interaction and uniaxial easy axis anisotropy is the key force in the formation of skyrmions.[21-26] The skyrmions in centrosymmetric materials are topologically equivalent to those in the chiral magnets[21-26] but possess two degrees of freedom, i.e. helicity and vorticity.[5] Since their internal degree of freedom is similar to that in the topologically trivial magnetic bubbles, they are also called skyrmionic bubbles.[24] The most interesting physics in skyrmionic bubbles is that the vorticity of their spin textures varies with the internal structure of the Bloch lines (BLs), resulting in a variety of spin textures.[22-26] For instance, a new type of spin texture formed by two skyrmions with opposite spin vorticity (the topological number equals 2), called a biskyrmion, has been experimentally discovered in the centrosymmetric tetragonal magnetite $La_{1-x}Sr_xMnO_3(x=0.315)$[23] and hexagonal $(Mn_{1-x}Ni_x)_{65}Ga_{35}(x=0.5)$[25] at temperatures around 60K and 300K, respectively. More recently, Yu *et al*.[24] have found a variety of spin textures in magnetic skyrmionic bubbles in orthorhombic magnetite $La_{1-x}Sr_xMnO_3(x=0.175)$ at 100K. Thus, the multifarious topological nature of skyrmionic bubbles offers us an opportunity to manipulate their topological spin textures through external stimuli.

Recently, the domain structures in the frustrated magnet with tunable UMA were studied numerically by Leonov *et al*[27]. They showed that the UMA strongly affects spin ordering. They also predicted that different spin structures, including isolated magnetic skyrmions, may coexist in a frustrated magnet, and that the isolated magnetic skyrmions also possess additional degrees of freedom (spin vorticity and helicity), similar to that in the magnetic skyrmionic bubbles. Their predictions are not only extremely interesting, but also point to further investigation of the variable topological spin textures in the frustrated magnets. Furthermore, Pereiro *et al*.[28] have theoretically shown that Heisenberg and DMI interactions in kagome magnets can overcome the thermal fluctuation and stabilize the skyrmions at relatively high temperatures, even at room temperature. Based on the theoretical investigations above, we revisit the frustrated magnets with kagome lattice to search for the possible skyrmionic bubbles that are able to stabilize at room temperature.

One of the promising materials is $Fe_3Sn_2$, suggested by Pereiro *et al.*,[28] which has a layered rhombohedral structure with alternate stacking of the Sn layer and the Fe-Sn bilayers along the *c*-axis, as shown in **Figure 1a**. The Fe atoms form bilayers of offset kagome networks, with Sn atoms throughout the kagome layers as well as between the kagome bilayers. Very importantly, this material is a non-collinear frustrated ferromagnet with a high Curie temperature $T_c$ of 640K, and shows a spin reorientation that the easy axis rotates gradually from the *c*-axis to the *ab*-plane as the temperature decreases.[29-32] Recently, a large anomalous Hall effect was observed in this material, which is strongly related to the frustrated kagome bilayer of Fe atoms.[33,34] In this communication, we report that magnetic skyrmionic bubbles with various spin textures can indeed be realized in the single crystals of $Fe_3Sn_2$ at room temperature. The emergence of skyrmionic bubbles and the magnetization dynamics associated with the transition of different bubbles via the field-driven motion of the Bloch lines are revealed by in-situ Lorentz transmission electron microscopy (LTEM), and further supported by the micromagnetic simulations and magnetic transport measurements. These results demonstrate that $Fe_3Sn_2$ facilitates a unique magnetic control of topological spin textures at room temperature, making it a promising candidate for further skyrmion-based spintronic devices.

High-quality single crystals of $Fe_3Sn_2$ were synthesized by a Sn-flux method, as described in the Methods section. These crystals are layered and exhibit mirror-like hexagonal faces with a small thickness (see **Figure S1**, SI). By using single-crystal X-ray diffraction (SXRD), the crystal parameters were identified as $a = b = 5.3074$Å and $c = 19.7011$Å, with respect to the rhombohedral unit cell (space group *R*-3m), agreeing well with the previous studies.[32-34] Having determined the unit lattice parameters and orientation matrix, we then found that the hexagonal face was normal to [001] with the (100), (010), and (110) faces around (see **Figure S2,** SI). In addition, as shown in **Figure S3**, both the temperature-dependent in-plane and out-of-plane magnetization curves measured on the bulk crystal indicate that the ferromagnetic transition temperature $T_c$ is about 640K.

**Figure 1b** shows the temperature-dependent UMA coefficient ($K_u$), which was estimated by the approximation of $K_u = H_k M_s/2$, where $M_s$ is the saturation magnetization and $H_k$ is the anisotropy field defined as the critical field above which the difference in magnetization between the two magnetic field directions (H//*ab* and H//*c*) becomes smaller than 2% (see **Figure S4**, SI). One can notice that the value of Ku increases monotonically with the decrease of temperature. Simultaneously, the Fe moments gradually rotate from the c-axis towards the ab-plane (see the inset of **Figure 1b**), demonstrating a gradual transformation from the magnetically easy axis (uniaxial magnetic anisotropy) into a magnetically easy plane with decreasing the temperature. The most important message conveyed to us by **Figure 1b** is that the magnetic domain configuration in $Fe_3Sn_2$ may vary over a very wide temperature range of 80-423K, because the domain structure is strongly affected by the magnetic anisotropy. Therefore, $Fe_3Sn_2$ should be a good platform for us to explore the correlation between the spin texture and Ku in a very wide temperature range.

We then imaged the magnetic domain structures using LTEM under zero magnetic field in the temperature range of 300K~130K, as shown in **Figure 1c-f**. The corresponding selected-area electron diffraction (SAED) patterns suggest that the sample is normal to the [001] direction (see the inset of **Figure 1b**). Nanosized stripe domains with an average periodicity of ~150nm were clearly observed. Notably, the value of periodicity is comparable to that in the bulk $(Mn_{1-x}Ni_x)_{65}Ga_{35}(x=0.5)$,[25] but nearly two times larger than that in the $La_{1-x}Sr_xMnO_3(x=0.175)$.[24] The sharp contrast between the dark stripe domains and the bright walls suggests that the domains possess out-of-plane magnetizations and are separated by Bloch domain walls. With the decrease of temperature, the stripes' periodicity *l* widened while the domain wall thickness *D* remained almost a constant (see **Figure S5**, SI). Interestingly, we found that when the temperature was lower than 130K, the stripe domains disappeared and vortex domains formed, indicating that the spin starts to lie into the *ab*-plane below 130K. The critical temperature of the LTEM sample coincides with that of the bulk sample, but with a slight deviation.[32-34] This feature suggests that the spin texture of domains in the bulk and LTEM samples show little differences. That can be attributed to the fact that the magnetic anisotropy in $Fe_3Sn_2$ is high enough to overcome the spin rearrangement effect resulting from the increase of demagnetizing energy in the thin LTEM sample. To understand the physics behind this domain structure transformation, we performed numerical simulations with estimated parameters of exchange constant (*A*) and anisotropy energy ($K_{u\perp}$) associated with the perpendicular component of the anisotropy field (see Methods). As shown in **Figure 1g**, the stripe domain gradually transformed into a vortex domain with decreasing $K_{u\perp}$, agreeing well with the LTEM images. This feature suggests that the domain morphology in $Fe_3Sn_2$ is mainly governed by the anisotropy perpendicular to the *ab*-plane, which is consistent with the simulated results based on a frustrated magnet.[27] It is well known that the magnetic domain structure also depends greatly on an external magnetic field. Therefore, we have simulated the domain structure under different magnetic fields that are perpendicular to the *ab*-plane (as shown in **Figure 1h**). It is interesting to note that the stripe domains gradually transformed into bubbles with increasing the external field. More strikingly, isolated magnetic skyrmions formed when the magnetic field increased to 400 mT.

To experimentally observe the domain structure variation with an external magnetic field predicted by the simulation, we imaged the domain structure under different magnetic fields at room temperature using LTEM. **Figure 2a-d** shows the over-focused LTEM images under different out-of-plane magnetic fields (the corresponding zero-field image is shown in **Figure S6**, SI). The gradual transformation from a stripe domain structure into magnetic skyrmionic bubbles is clearly observed as the magnetic field increases from 0 to 860mT. In **Figure 2a**, we show a snapshot of the transformation process under a magnetic field of 300mT. One can notice that the stripe domains, dumbbell-shaped domains and magnetic bubbles coexist in the image. We actually observed that, during the evolution of the domain structures, the stripe domains gradually broke into the dumbbell-like domains first before evolving into the magnetic bubbles. When the magnetic field increased above 800mT, all the stripes and dumbbell-like domains completely transformed into magnetic bubbles. One should note that we only changed the external magnetic field and kept all other conditions constant in **Fig. 2a-d**. Therefore, the changes in domain structure can be entirely ascribed to the external field effect. A closer analysis of the entire transformation process reveals that the spin texture of the magnetic bubbles changes dramatically with increasing the magnetic field, as shown in **Figure 2e-h** (the bubbles with different spin textures are notated by different numbers). The structural evolution of the magnetic bubbles with increasing external magnetic field should be closely related to the change of topology in the bubbles, as previously observed.[24,26]

To characterize the topological spin textures of the magnetic bubbles, a transport-of-intensity equation (TIE) was employed to analyze the over- and under-focused LTEM images. **Figure 2i-l** displays the spin textures of the bubble domains shown in **Figure 2e-h**. The white arrows show the directions of the in-plane magnetic inductions, while the black regions represent the domains with out-of-plane magnetic inductions. Bubble "1", composed of a pair of open Bloch lines (BLs), is characteristic of the domain structure commonly observed in ferromagnetic compounds with uniaxial magnetic anisotropy. In this bubble, the topological number *N* is determined to be 0, because the magnetizations of the BLs are nonconvergent. Further increase of the magnetic field induced the formation of bubble "2", which has two arc-shaped walls with opposite helicity, separated by two BLs. Similar to that of bubble "1", the spin texture of bubble "2" is also not convergent, leading to the same topological number, N=0. However, when the magnetic field increased above 800mT, the topological number of bubbles "3" and "4" transforms from 0 to 1, being equal to that of skyrmions. As shown in **Figure 2k**, bubble "3" has a pendulum structure with two BLs, in which the spin textures become convergent. Compared with the recent results obtained in $La_{1-x}Sr_xMnO_3$(x=0.175) by Yu *et al*,[24] bubble "3" can also be regarded as a specific type of rarely observed skyrmionic bubble. Bubble "4" is the most orthodox skyrmionic bubble, possessing the same domain structure as the skyrmions observed in chiral magnets. One important feature of this bubble is that the thickness of the Bloch wall is comparable to the radius of the bubble, leading to a small core region. Following previous reports,[24] we understand that bubble "3" transformed into bubble "4" through the motion of BLs driven by the magnetic field. Therefore, although the magnetic textures in bubbles "3" and "4" are strikingly different, they are homeomorphic in topology. To understand the above transformations of spin texture in more detail, we successfully recorded the transformation process from the topologically nontrivial magnetic bubbles to the topologically protected skyrmions in a $Fe_3Sn_2$ (001) thin-plate using in-situ LTEM (**see Supplementary Movie**, SI). **Figure 3a-f** presents several snapshots of the transformation process, selected from a movie taken by LTEM at 300 K and under different out-of-plane magnetic fields. We observed a pair of BLs move along the bubble wall and eventually die out within 8.3 seconds, as the field increased from 840mT to 850mT. These results demonstrate clearly that isolated magnetic skyrmions can be realized in the frustrated $Fe_3Sn_2$ magnet through BL motion by tuning the external magnetic field, even at room temperature.

In addition to the observation of various spin textures of skyrmionic bubbles during BL motion, we further found that the magnetic domain configurations after turning off the external fields depend strongly on the strength of the external fields. If the thin-plate sample is first magnetized to saturation (i.e. the sample is in single-domain state), then the domain will return to the multi-stripe state after turning off the external magnetic field (see **Figure S7**, SI). However, if the sample is magnetized to an intermediate state (for example, at a field of 700mT as shown in **Figure S8**, SI), the domain structure evolves differently after turning off the external magnetic field. **Figure 4a** shows the under-focused LTEM image taken at 300K after turning off the external field of 700mT by which the sample was magnetized to an intermediate state. The coexistence of different types of magnetic domains, e.g. magnetic bubbles, stripes and dumbbell-like domains is clearly observed. After careful analysis of the image using TIE, the detailed spin texture of the domain enclosed by the square in **Figure 4a** is shown in **Figure 4e**. Unexpectedly, the magnetic bubbles possess three concentric rings. It is particularly interesting that the winding directions of the inner and outer rings are opposite to that of the middle ring, indicating that the helicity reverses inside the bubbles. These spontaneous bubbles can be considered skyrmions, analogous to those observed in $BaFe_{12-x-0.05}Sc_xMg_{0.05}O_{19}$.[22]

To explore the evolution of the domain structure with temperature, the domain structures in the same region as shown in **Figure 4a** were also imaged at several lower temperatures without changing the external field (**Figure 4b-d**). These results help us to understand how the domain structure evolves under the influence of different magnetic energies, i.e. magnetic anisotropy, exchange energy, and static magnetic energy. As the temperature decreases, the size of the spontaneous bubbles change slightly and the stripes gradually transform into skyrmions. Consequently, the number of spontaneous bubbles increases with decreasing temperature. The maximum density of the bubbles appears at 250K. The spontaneous bubbles vanish as the temperature further decrease below 100K, due to the transformation from uniaxial, out-of-plane anisotropy into in-plane anisotropy (**Figure 1b**). The corresponding spin textures of the skyrmions marked in **Figure 4b-d** were also extracted by using TIE analysis, as shown in **Figure 4f-h**. It was found that the size of the innermost ring increases with decreasing the temperature. Micromagnetic simulations are currently ongoing in order to explore the interplay among the different magnetic parameters behind these features.

The formation of magnetic skyrmionic bubbles and isolated skyrmion spin textures in the bulk $Fe_3Sn_2$ single crystals were further studied and confirmed by magnetic and magneto-transport measurements, similar to previous studies of other skyrmion-based materials.[7,20,25,35] It should be noted here that the sample for magnetic and magnetotransport measurements is from the same crystals used for LTEM observations. In **Figure 5a**, we show the dependence of magnetoresistance (MR) on the magnetic field ($H$) that is applied normal to the *ab*-plane in the temperature range of 100-400K. The inset shows the details of the MR-$H$ curve at 300K, in which two broad peaks are clearly seen at about 200mT and 800mT, respectively. Based on the LTEM results shown in **Figure 2** and our analysis, it can be concluded that the peak at 200mT ($H_a$) reflects the starting point of the transformation from stripes to magnetic bubbles, whereas the peak at 800mT ($H_m$) represents the starting point of the transformation from magnetic bubbles to insulated skyrmions. When the magnetic field increases above 900mT ($H_c$), the sample reaches the magnetically saturated state (see the *M-H* curves in **Figure S4**) in which the spins are aligned along the field direction, and consequently all the skyrmions die out. A close inspection of the MR-H curves in the range of 400-100K reveals that all three critical fields shift monotonically with temperature as indicated by the dotted lines. As shown in **Figure 5b**, we can also identify the three critical fields in the field-dependent AC-susceptibility, though the critical fields are slightly lower than those observed in the MR results and show slightly weaker temperature dependence. Based on the LTEM, MR, and AC-susceptibility results, we can roughly create a magnetic phase diagram as depicted in **Figure 5c**. Based on the phase diagram, we predict that the topological spin texture states in $Fe_3Sn_2$ may extend to a much higher temperature, perhaps up to Curie temperature (~640K). However, due to the technical limitations of our current measurements, we could not perform the experiments at a temperature higher than 400K. It is well known that stable skyrmionic states at high temperatures are critical for technical applications in magnetic storage and spintronics devices. Therefore, the observation of a skyrmion state in $Fe_3Sn_2$, not only in a very wide temperature range but also at a high temperature, strongly suggests that $Fe_3Sn_2$ is a very promising material for both skyrmion physics and potential technical applications of magnetic skyrmions.

The ongoing and future studies will include electrically probing various exciting phenomena in this material, such as the skyrmion Hall effect,[36, 37] and quantizing the transport of magnetic skyrmions,[38,39] similar to the study of emergent electrodynamics of skyrmions in bulk chiral materials.[40]

**Experimental Section**

***Sample Preparation*:** Single crystals of $Fe_3Sn_2$ were synthesized by using the Sn-flux method with a molar ratio of Fe : Sn = 1:19. The starting materials were mixed together and placed in an aluminum crucible with higher melting temperatures at the bottom. This process was performed in a glove box filled with argon gas. To avoid the influence of volatilization of Sn at high temperatures, the whole assembly was first sealed inside a tantalum (Ta) tube under proper Ar pressure. The Ta tube was then sealed in a quartz tube filled with 2 mbar Ar pressure. The crystal growth was carried out in a furnace by heating the tube from room temperature up to 1150℃ over a period of 15 h, holding at this temperature for 72 h, cooling to 910℃ over 6 h, and subsequently cooling to 800℃ at a rate of 1.5 K/h. The excess Sn flux was removed by spinning the tube in a centrifuge at 800℃. After the centrifugation process, most of the flux contamination was removed from the surfaces of crystals and the remaining flux was polished.

***Magnetic and Transport Measurements:*** The magnetic moment was measured by using a Quantum Design physical properties measurement system (PPMS) between 10K and 400K, whereas the magnetic moment above 400K was measured by using a vibrating sample magnetometer (VSM). To measure the (magneto-) transport properties, several single crystals were milled into a bar shape with a typical size of about 0.6 × 0.4 × 0.05 mm$^3$. Both longitudinal and Hall resistivity were measured using a standard four-probe method on a Quantum Design PPMS. The field dependence of the Hall resistivity was obtained after subtracting the longitudinal resistivity component.

*LTEM Measurements*: The thin plates for Lorentz TEM observations were cut from bulk single-crystalline samples and thinned by mechanical polishing and argon-ion milling. The magnetic domain contrast was observed by using Tecnai F20 in the Lorentz TEM mode and a JEOL-dedicated Lorentz TEM, both equipped with liquid-nitrogen, low-temperature holders (≈100 K) to study the temperature dependence of the magnetic domains. The magnetic structures were imaged directly in the electron microscope. To determine the spin helicity of the skyrmions, three sets of images with under-, over-, and just (or zero) focal lengths were recorded by a charge-coupled device (CCD) camera, and then the high-resolution in-plane magnetic induction distribution mapping was obtained by QPt software, based on the transport of the intensity equation (TIE). The inversion of magnetic contrast can be seen by comparing the under- and over-focused images. The colors and arrows depict the magnitude and orientation of the in-plane magnetic induction. The objective lens was turned off when the sample holder was inserted, and the perpendicular magnetic field was applied to the stripe domains by increasing the objective lens gradually in very small increments. The specimens for the TEM observations were prepared by polishing, dimpling, and subsequently ion milling. The crystalline orientation of the crystals was determined by selected-area electron diffraction (SAED).

*Micromagnetic Simulations*: Micromagnetic simulations were carried out with three-dimensional object oriented micromagnetic frame work (OOMMF) code, based on the LLG function.[41] Slab geometries of dimensions 2000 nm × 2000 nm × 100 nm were used, with rectangle mesh of size 10nm×10nm×10nm. We used a damping constant α=1 to ensure quick relaxation to the equilibrium state. The material parameters were chosen according to the experimental values of Fe$_3$Sn$_2$, where the saturation magnetization $M_s$ = 5.66×10$^5$ A/m at room temperature, and the uniaxial magnetocrystalline anisotropy constant $K_u$=1.8×10$^5$ J/m$^3$. As the magnetic moment aligned obliquely along the *c*-axis, we defined $K_{u\perp}$ as the anisotropy energy associated with the perpendicular component of the anisotropic field, i.e. $K_{u\perp} = \left(\frac{1}{2}\right) H_{k\perp} M_s$. The exchange constant *A* was estimated to be 1.4×10$^{-11}$ A/m by $D = \pi\sqrt{A/K_u}$, where *D* is the domain wall width obtained from the LTEM results. These three parameters vary with temperature, hence we investigated the dependence of domain morphology on exchange energy (*A*) and anisotropy energy as shown in Figure 1g by fixing the $M_s$. The equilibrium states are all obtained by fully relaxing the randomly distributed magnetization. The simulations on the field-dependent domain structures at 300K (as shown in **Figure 1h**) were conducted at zero temperature (no stochastic field) but the values of parameters correspond to 300K, because the change tendency of magnetic domain alters little by temperature.

**Supporting Information**

Supporting Information is available from the Wiley Online Library or from the author.

**Acknowledgements**

The authors thank Jie Cui and Dr. Yuan Yao for discussions and their help in LTEM experiments. This work is supported by the National Natural Science Foundation of China (Grant Nos. 11474343, 11574374, 11604148, 51471183, 51590880, 51331006 and 5161192), King Abdullah University of Science and Technology (KAUST) Office of Sponsored Research (OSR) under Award No: CRF-2015-2549-CRG4, China Postdoctoral Science Foundation NO. Y6BK011M51, a project of the Chinese Academy of Sciences with grant number KJZD-EW-M05-3, and Strategic Priority Research Program B of the Chinese Academy of Sciences under the grant No. XDB07010300.

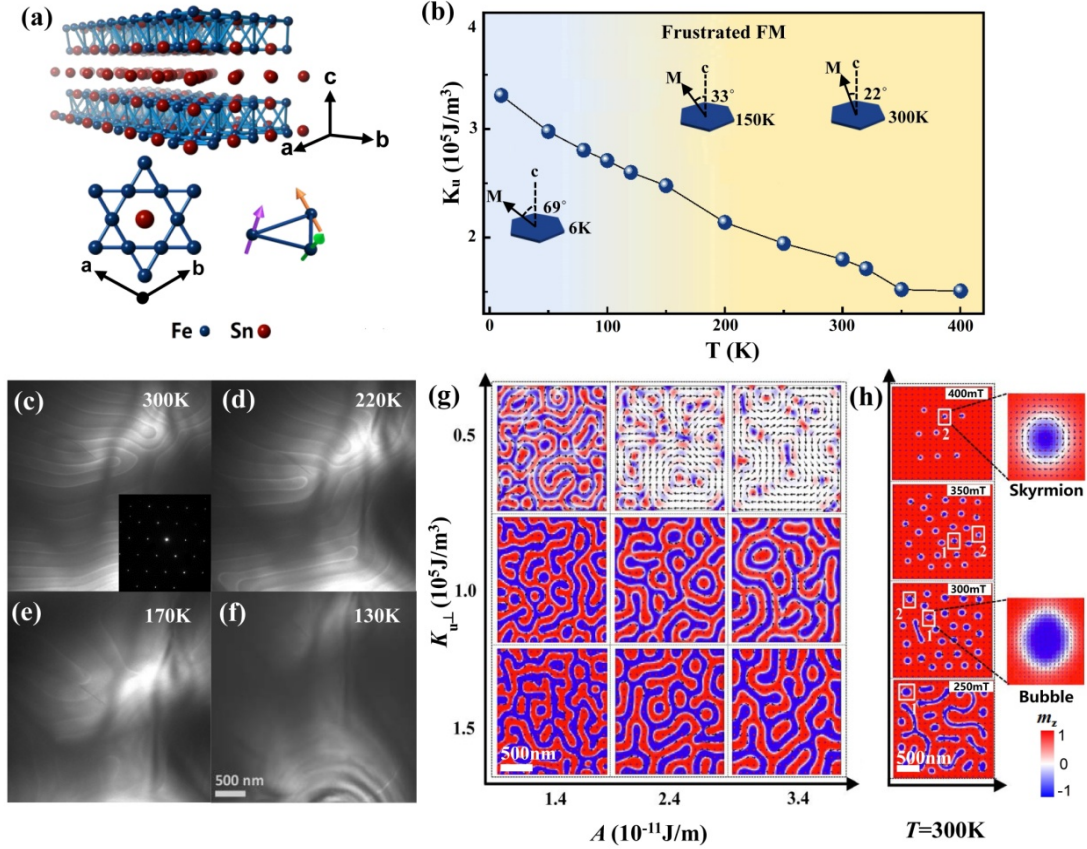

**Figure 1.** Structure, magnetic properties, and micromagnetic stimulations of $Fe_3Sn_2$. **a)** The crystal structure of $Fe_3Sn_2$ (up), a top view of the kagome layer of Fe atoms (down, left) and a possible spin (arrows) configuration of the Fe atoms (down, right). **b)** The temperature dependence of anisotropy constant ($K_u$) in the temperature range 10-400K. The insets (from right to left) show the schematic of the angle between the magnetic easy axis and the *c*-axis at 300, 150, and 6K, respectively. **c-f)** The representative images of the domain structures of same area in a $Fe_3Sn_2$ thin-plate taken by Lorentz transmission electron microscopy (LTEM) with an electron beam perpendicular to the *ab*-plane of $Fe_3Sn_2$ when the sample temperature was lowered from 300 K to 130K in zero external magnetic field. The **inset** of (c) shows the corresponding selected-area electron diffraction (SAED) pattern. **g)** The plan view of equilibrium states under different ratios of exchange constant *A* and perpendicular component for the magnetocrystalline magnitude $K_{u\perp}$ for a fixed $M_s=5.7\times10^5$ A/m and thickness (100nm) in zero external magnetic field. The magnetization along the z-axis ($m_z$) is represented by regions in red ($+m_z$) and blue ($-m_z$), whereas the in-plane magnetization ($m_x$, $m_y$) is represented by the white regions. **h)** The simulated field-dependent domain morphology under several magnetic fields that capture the domain evolution from stripes to type-II bubbles and skyrmionic bubbles.

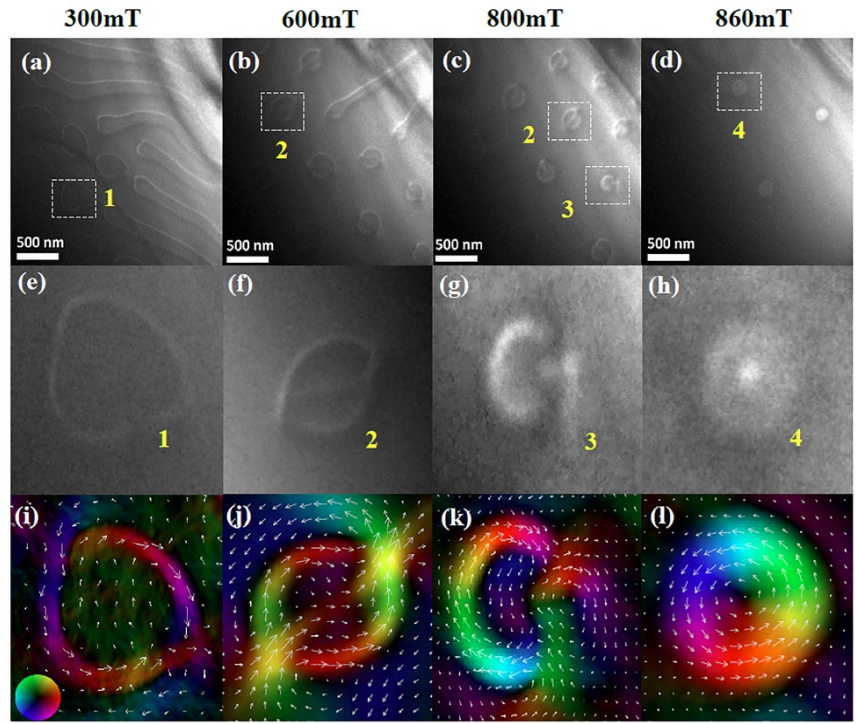

**Figure 2.** Magnetic field dependence of the magnetic domain morphology imaged using LTEM at 300K. **a-d)** The over-focused LTEM images under different out-of-plane magnetic fields at 300K. The regions in the white boxes show the different types of magnetic bubble domains. **e-h)** Enlarged LTEM images of the white boxes in (a-d) showing the magnetic bubble domains. **i-l)** Corresponding spin textures for the bubble domains shown in (e-h), extracted from the analysis using TIE. Colors (the **inset** of (i) shows the color wheel) and white arrows represent the direction of in-plane magnetic induction, respectively, whereas the dark color represents the magnetic induction along the out-of-plane direction. **(e)** and **(f)** display the type-II bubbles, and **(k)** and **(l)** show different types of skyrmionic bubbles.

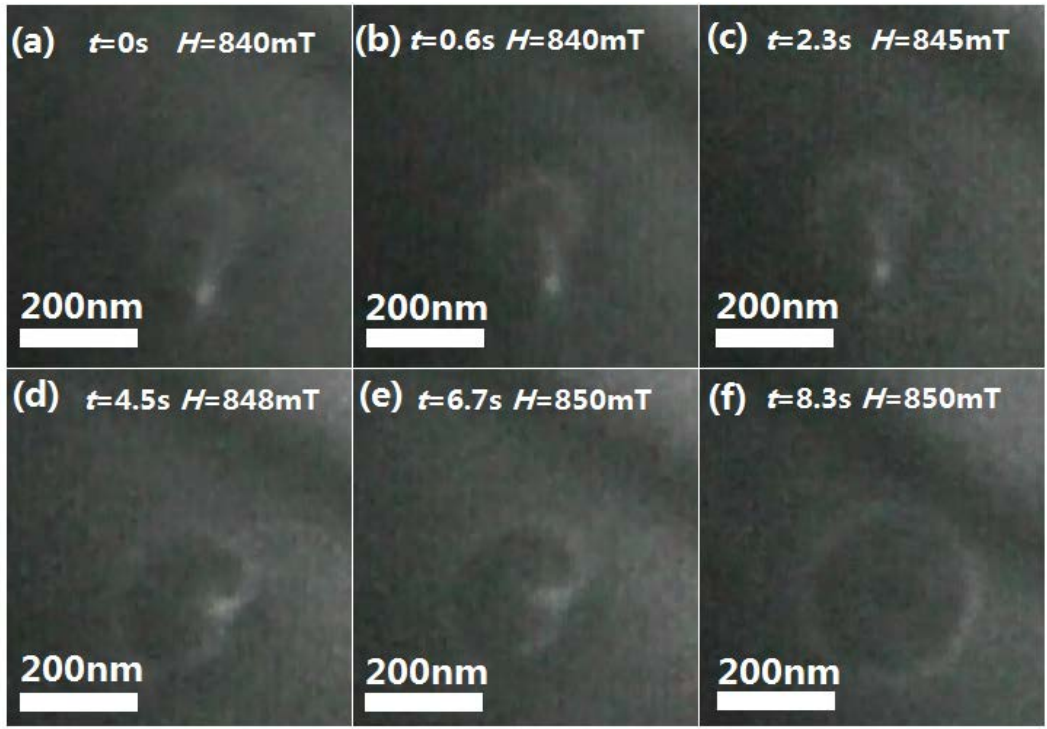

**Figure 3.** Evolution of the magnetic bubble through Bloch line (BL) motion, induced by the magnetic field. **a-f)** Series of LTEM images of bubble "3" observed at different times and fields applied along the *c*-axis. The field was increased slowly for 8 seconds from 840 to 850mT.

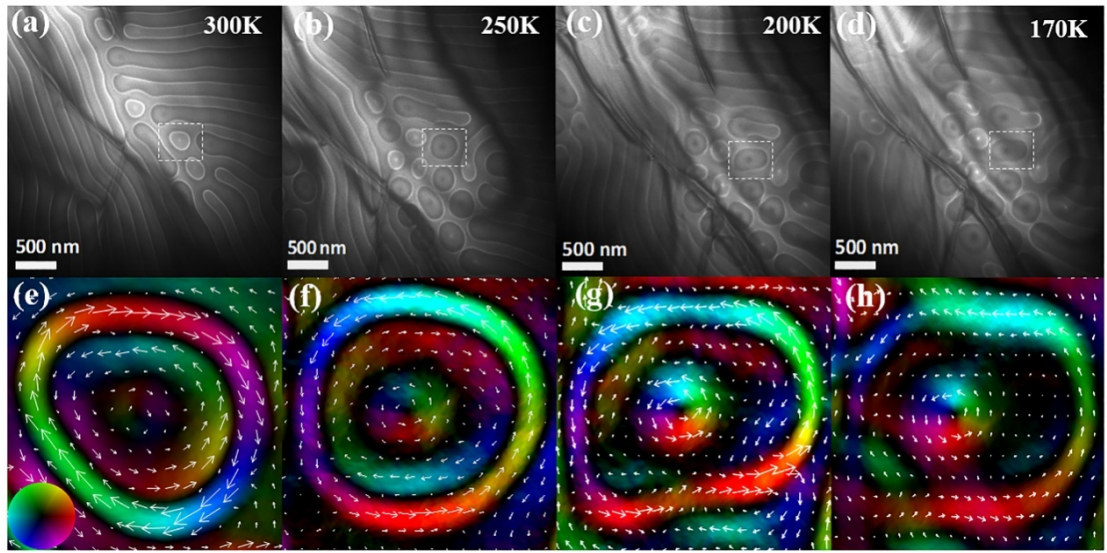

**Figure 4.** LTEM images of the magnetic domain structures taken at different temperatures after turning off the external magnetic field applied along *c*-axis. The field of 700mT is not strong enough to saturate the sample magnetically. **a-d)** Stripe domains and skyrmionic bubbles coexist in the sample over the temperature range of 300K to 170K, after turning off the magnetic field. The regions enclosed in the squares are the skyrmionic bubbles. **e-h)** The corresponding magnetization textures obtained from the TIE analysis for the bubble domains in the squares in (a-d). The **inset** of (e) shows the color wheel. The spontaneous skyrmionic bubbles with triple-ring structures and random helicities are observed.

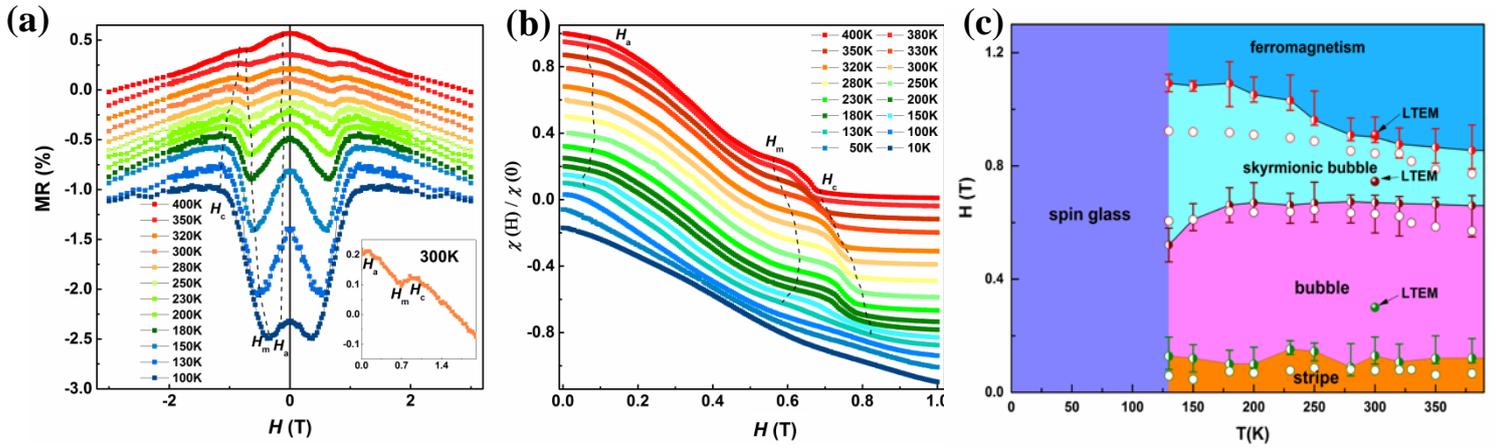

**Figure 5.** Field dependence of magnetoresistance (MR), AC-susceptibility and the magnetic phase diagrams of Fe$_3$Sn$_2$. **a)** The magnetic field ($H$) dependence of MR obtained in the temperature range of 100K-400K, with magnetic fields applied normal to the *ab*-plane. The inset shows the details of the MR-$H$ curve measured at 300K. **b)** The magnetic field dependence of the real part of AC-susceptibility $\chi$ in the temperature range of 100K-400K. **c)** Magnetic phase diagram of bulk sample in the magnetic field versus the temperature plane, as deduced from the temperature dependence of MR and $\chi$ curves. The three fully filled circles indicate the experimental data obtained from the in situ LTEM observations. The error bars were added based on the results measured on three different samples.

**Various and spontaneous magnetic skyrmionic bubbles** are experimentally observed for the first time, at room temperature in a frustrated kagome magnet $Fe_3Sn_2$ with unixial magnetic anisotropy. The magnetization dynamics were investigated using in-situ Lorentz transmission electron microscopy, revealing that the transformation between different magnetic bubbles and domains are via the motion of Bloch lines driven by applied external magnetic field. The results demonstrate that $Fe_3Sn_2$ facilitates a unique magnetic control of topological spin textures at room temperature, making it a promising candidate for further skyrmion-based spintronic devices.

Keywords: Skyrmionic bubbles; Topological spin textures; Kagome magnet; $Fe_3Sn_2$; Spintronic devices

*Zhipeng Hou\*, Weijin Ren\*, Bei Ding\*, Guizhou Xu, Yue Wang, Bing Yang, Qiang Zhang, Ying Zhang, Enke Liu, Feng Xu, Wenhong Wang, Guangheng Wu, Xi-xiang Zhang, Baogen Shen, Zhidong Zhang*

Observation of Multiple and Spontaneous Skyrmionic Magnetic Bubbles at Room Temperature in a Frustrated Kagome Magnet with Uniaxial Magnetic Anisotropy

TOC Figure

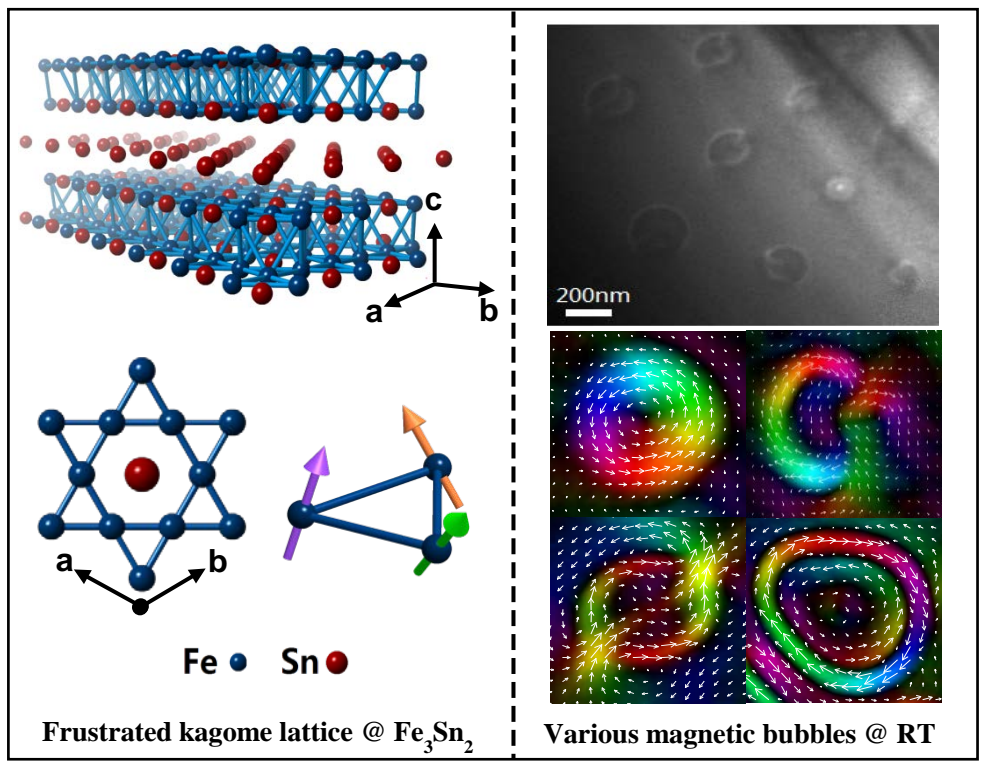

**Frustrated kagome lattice @ $Fe_3Sn_2$**   **Various magnetic bubbles @ RT**

## Supplementary Information to

**Observation of Various and Spontaneous Magnetic Skyrmionic Bubbles at Room-temperature in a Frustrated Kagome Magnet with Uniaxial Magnetic Anisotropy**

*Zhipeng Hou\*, Weijin Ren\*, Bei Ding\*, Guizhou Xu, Yue Wang, Bing Yang, Qiang Zhang, Ying Zhang, Enke Liu, Feng Xu, Wenhong Wang, Guangheng Wu, Xi-xiang Zhang, Baogen Shen, Zhidong Zhang*

Single-crystal X-ray diffraction (SXRD) was performed on the crystal shown in **Figure S1** with a Bruker APEX II diffractometer using Mo K-alpha radiation (lambda = 0.71073 A) at room temperature. Exposure time was 10 seconds with a detector distance of 60 mm. Unit cell refinement and data integration were performed with Bruker APEX3 software. A total of 180 frames were collected over a total exposure time of 2.5 hours. The crystal lattice parameters were established to be $a = b = 5.3074$Å, $c = 19.7011$Å with respect to the rhombohedral unit cell (space group *R*-3m), agreeing well with the previous studies. As shown in **Figure S2**, to ascertain the crystal orientations, the sample was mounted on a holder and Bruker APEX II software was used to indicate the face normal of the crystal, after the unit cell and orientation matrix were determined. One can notice that the hexagonal face is normal to [001] with the (100), (010), and (110) faces around.

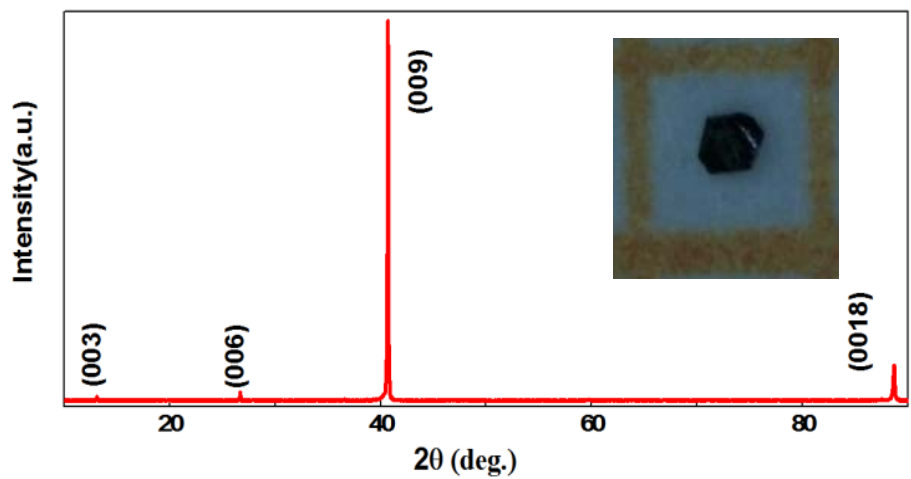

**FigureS1.** X-ray diffraction pattern of a $Fe_3Sn_2$ single crystal along the perpendicular direction of hexagonal surface, which indicates that the hexagonal surface is parallel to the *ab*-plane and perpendicular to the *c*-axis. Inset: The typical photograph of $Fe_3Sn_2$ single crystal placed on a millimeter grid. The crystal is 0.3 mm × 0.3 mm × 0.2 mm in size and possesses hexagonal mirror-like surfaces.

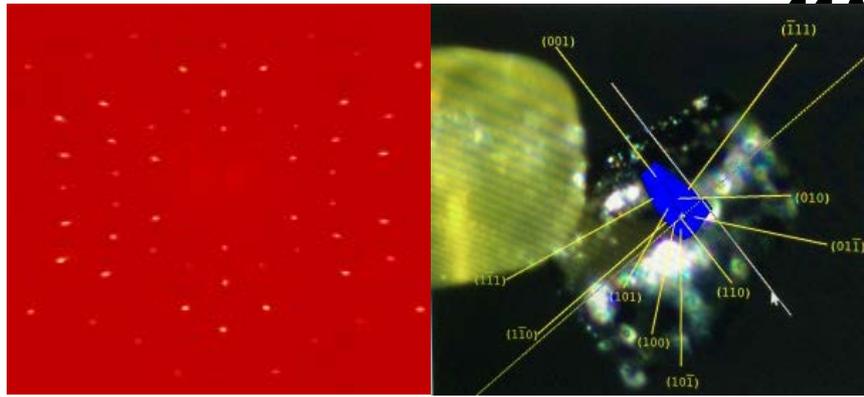

**Figure S2.** (a) Single-crystal X-ray diffraction processing image of the (00$l$) plane in the reciprocal lattice of $Fe_3Sn_2$ obtained on the crystal mentioned above. No diffuse scattering is seen and all the resolved spots fit the crystal lattice structure established for $Fe_3Sn_2$. (b) Various crystal planes and their corresponding normal directions of $Fe_3Sn_2$ single crystal. The yellow striped region is the glue to affix the crystal to the holder.

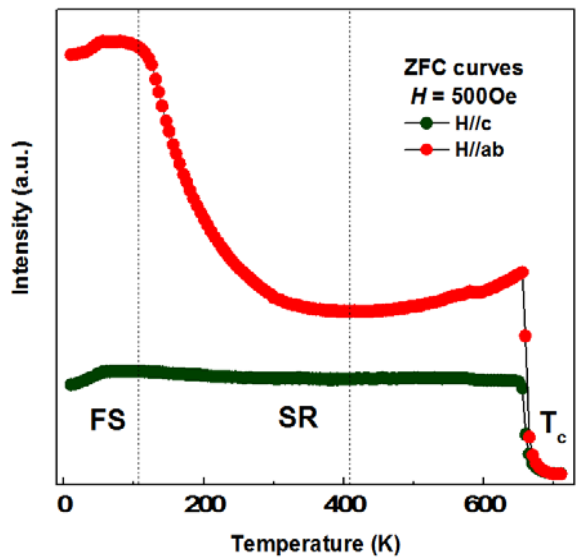

**Figure S3.** Temperature dependence of magnetization with the field-cooling (FC) model in an external magnetic field of 500Oe between 5K and 700K. As is shown in **Figure S3**, the Curie temperature $T_c$ is established to be 660K, which is similar to previous reports. When the temperature falls below $T_c$, the magnetization for both magnetic fields ($H//c$ and $H//ab$) first decreases and then starts to increase at 420K, reaching a maximum at 80K. When the temperature decreases below 80K, the slight decrease in magnetization can be attributed to the entrance of the spin glass state (SGS).

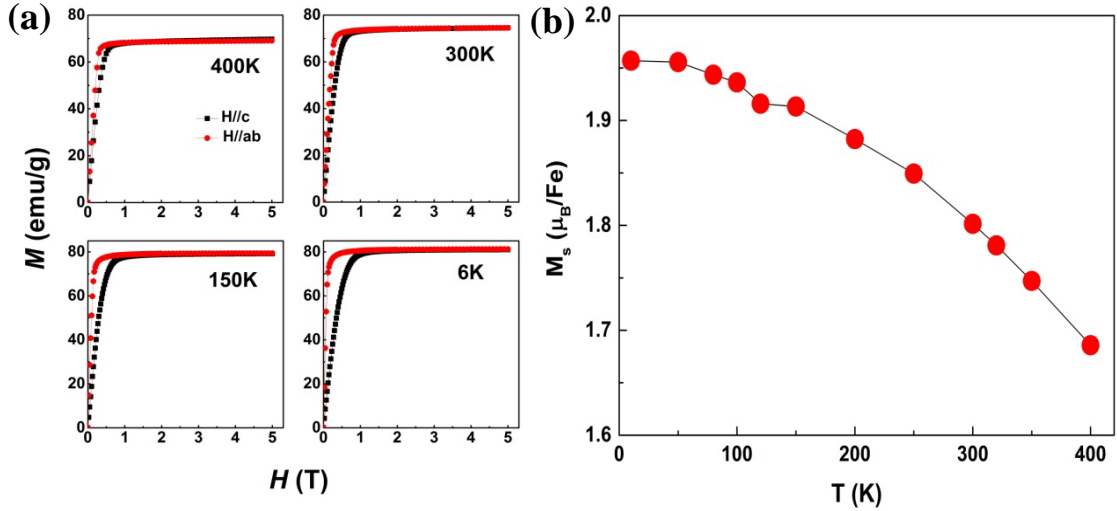

**Figure S4. a)** The magnetic field dependence of magnetization in fields parallel to the *c*-axis (black line and black symbol) and *ab*-plane (red line and red symbol) in the temperature range of 400K-6K. **b)** The temperature dependence of the saturation magnetization $M_s$.

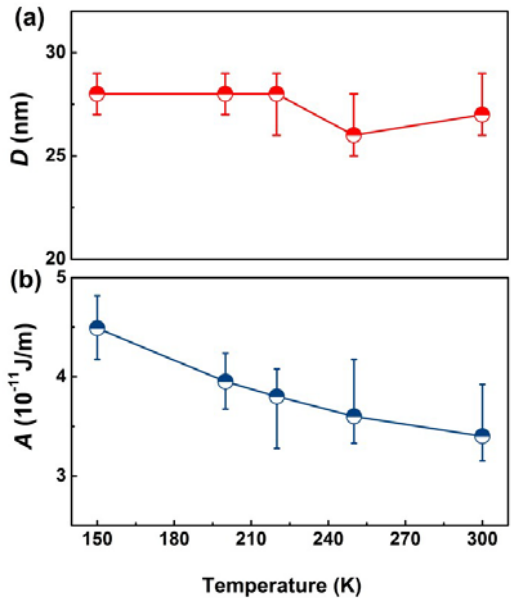

**FigureS5. a)** The temperature dependence of domain wall thickness *D*. The error bar denotes the deviation of three individual width measurements. One can notice that the value of *D* is nearly independent from the change in temperature. The exchange stiffness constant *A* can be established by using the equation, $A = \frac{DK_u^2}{\pi^2}$. **b)** The temperature dependence of *A*. By decreasing the temperature, the value of *A* increases correspondingly.

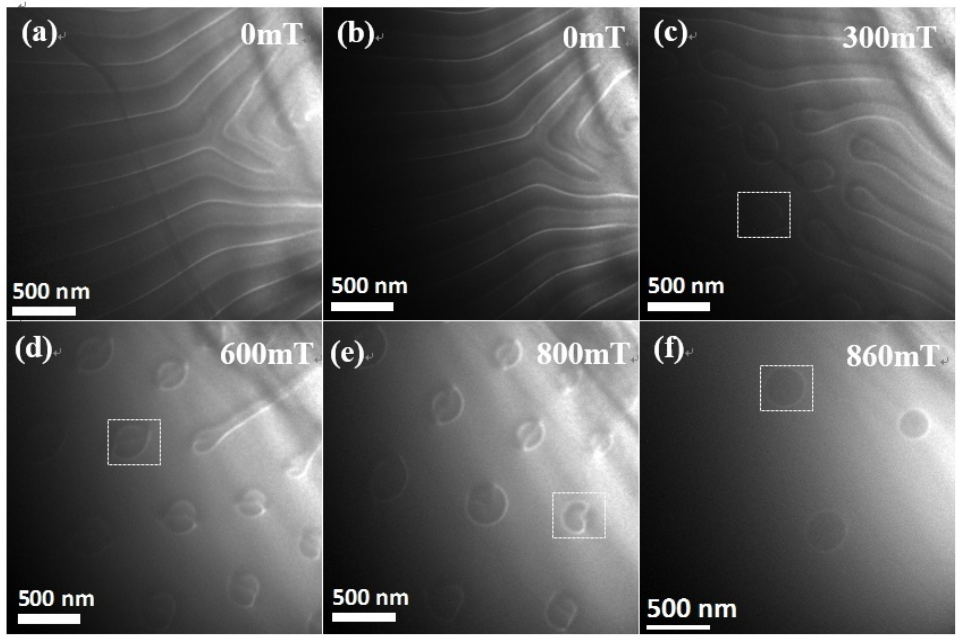

**FigureS6. a-b)** The over- and under-focused LTEM images under zero magnetic field at 300K. **c-f)** Corresponding under-focused TEM images for **Figure 2 (a-d)**. The boxed regions correspond to the magnetic bubbles shown in **Figure 2 (a-d)**.

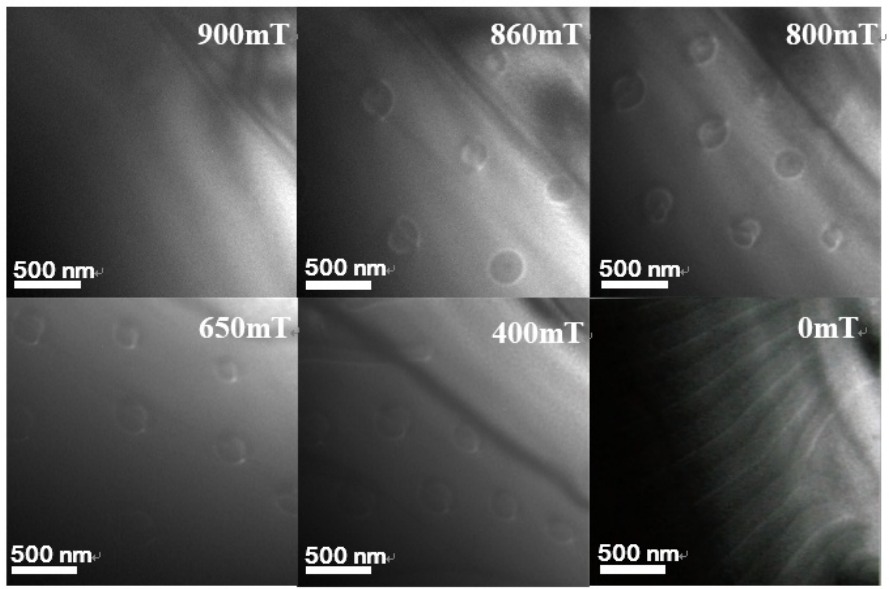

**Figure S7.** The under-focused LTEM images after a saturated magnetization. When the sample is magnetized to a saturated state, the domain reverts to the stripe.

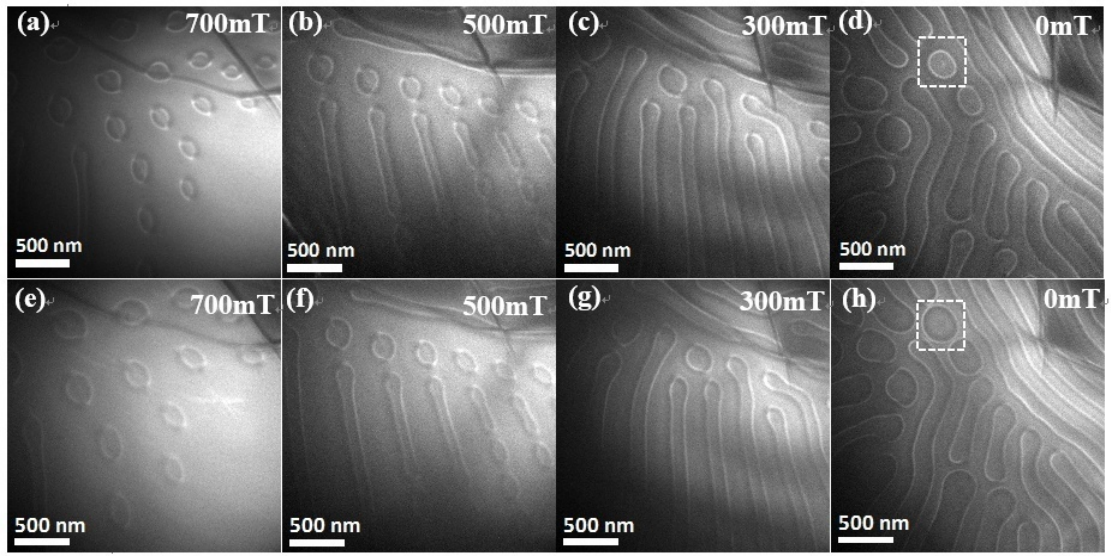

**Figure S8.** The corresponding over-focused (**a, b, c, d**) and under-focused (**e, f, g, h**) LTEM after an unsaturated magnetization in a different region from that in Figure 3. If the sample is magnetized to an intermediate state, then the skyrmionic bubbles with concentric rings appear after the magnetic field decreases to zero.